\documentclass{article}

\usepackage{PRIMEarxiv}

\usepackage[utf8]{inputenc} 
\usepackage[T1]{fontenc}    
\usepackage{hyperref}       
\usepackage{url}            
\usepackage{booktabs}       
\usepackage{amsfonts}       
\usepackage{nicefrac}       
\usepackage{microtype}      
\usepackage{lipsum}
\usepackage{graphicx}
\graphicspath{{media/}}     

\usepackage{setspace}
\usepackage{caption}
\usepackage{subcaption}
\usepackage{float}
\usepackage{gensymb}
\usepackage{hyperref}
\usepackage{xcolor}
\usepackage[version=3]{mhchem}
\usepackage{mwe,tikz}\usepackage[percent]{overpic}
\usepackage{graphicx,stackengine}
\usepackage{xspace}
\usepackage{fixltx2e}
\usepackage{amsmath}
\usepackage{amssymb}
\usepackage{enumerate}
\usepackage{marginnote}
\usepackage{siunitx}
\usepackage{longtable,tabularx}
\usepackage{booktabs}
\usepackage{makecell}
\usepackage{nomencl} 
\usepackage{wasysym}

\newcommand{\C}{$^\circ$C\xspace}

\title{Real-time quantitative imaging of RTV silicone pyrolysis

}

\author{
  Collin W. Foster\thanks{\textit{\underline{Corresponding author}}: 
  \textbf{collinf3@illinois.edu}} , Sreevishnu Oruganti, Francesco Panerai \\
  Center for Hypersonics and Entry Systems Studies, Department of Aerospace Engineering \\
  University of Illinois at Urbana-Champaign \\
  Urbana, 61801, Illinois, USA\\
  \texttt{\{collinf3, so24, fpanerai\}@illinois.edu} \\
}

\begin{document}
\maketitle

\begin{abstract}
Quantitative microstructural analysis of Room Temperature Vulcanized (RTV) silicone pyrolysis at high temperatures is presented. RTV is used as a bonding agent in multiple industries, particularly filling gaps in ablative tiles for hypersonic (re-)entry vehicles and fire prevention. Decomposition of RTV is resolved in real time using \textit{in situ} high-temperature X-ray computed micro-tomography. Full tomographies are acquired every 90~seconds for four different linear heating rates ranging from 7 to 54\C/min. The microstructure is resolved below 5 $\mu$m/pixel, allowing for a full quantitative analysis of the micro-structural evolution and porous network development. Results are highly heating rate dependent, and are evaluated for bulk sample volume change, porosity, pore network size, and observed densification from X-ray attenuation. The outcome of this work is critical to develop multi-physics models for thermal response.
\end{abstract}


\section{Introduction}
High temperature materials are advancing in the fields of nuclear reactors, fire-prevention, and thermal protection systems (TPS) for hypersonic flight. With the increasing applications of such materials, their characterization as a function of temperature is critical to application and modeling efforts. 
Room Temperature Vulcanizing Silicone (RTV) is a material system noted for its excellent thermal and mechanical properties and resistance to chemical degradation at high temperatures \cite{lee1977,cherney1999,fenimore1966}. 
RTV is also adaptable as a gap-filler for ablative heat shields, seeing widespread applications in exploration missions such as Mars2020 \cite{maddock2020}, Mars Science Laboratory (MSL) heatshield \cite{beck2009}, and other sample-return missions \cite{venkatapathy2009}. RTV's chemical composition contains iron compounds, which are well-known for being flame retardant, smoke suppressant, and can delay thermal decomposition, all of which benefit the fire-safety industries \cite{weil2003,chen2016}. 
Data recovered from the Space Shuttle missions and MSL describe the extreme aerothermal (re-)entry environments with temperatures over 2500\C where RTV was used as a bonding agent \cite{curry1993,white2013,hwang2016}. RTV and silicone-based coatings are unique in that they are also intumescent, meaning that they will swell when heated to create an insulating barrier from the source of the heat\cite{gardelle2013}. 
This property of intumescence is helpful in coating applications (such as on wood, plastic, or metal) \cite{halpern1984} but must be investigated further when swelling could cause unforeseen mechanical stress on a highly compact TPS \cite{mckeon1969,fu2018,fu2020}. RTV as a gap-filler on tiled TPS is observed to create protuberances in hypersonic flow as it ablates differently from the heat shield tiles, transitioning the flow from laminar to turbulent, resulting in areas of higher heating on the heatshield \cite{driver2014}. 
This behavior can be non-destructively examined with X-ray computed micro-tomography ($\mu$-CT), as a method to examine the inner-morphology of the material system as it pyrolyzes at high temperatures. Performing $\mu$-CT is a proven technique for evaluating complicated porous networks that are chemically decomposing, and the results can then be used to model multi-scale material behavior \cite{panerai2017, ferguson2018, semeraro2020, semeraro2021, roberts2014}. 
Recent work has thoroughly characterized the microstructure of RTV, emphasizing the utility of $\mu$-CT as a quantitative tool for examining the formation of voids at high temperatures \cite{oruganti2022, oruganti2023}. 
However, further investigation must be made into its heating rate dependence on micro-scale (1-\SI{100}{\micro\meter}) porosity evolution. 
Experiments to resolve the microstructural evolution of RTV at various heating ramp rates will be used to inform multi-physics codes, to then incorporated into larger TPS and fire modeling efforts \cite{meurisse2018,okyay2019,torresherrador2020,torresherrador2019,lachaud2017,lachaud2014}. 
Thus, this study utilizes synchrotron radiation to resolve the microstructural effects of pyrolysis at a range of heating rates  \emph{in situ}. This is quantified by measurements of porosity, total volume change, X-ray attenuation, followed by further discussion with scanning electron microscopy (SEM) images. 

\section{Materials and Methods}
\begin{figure}[H]
	\centering
    \includegraphics[width=\textwidth]{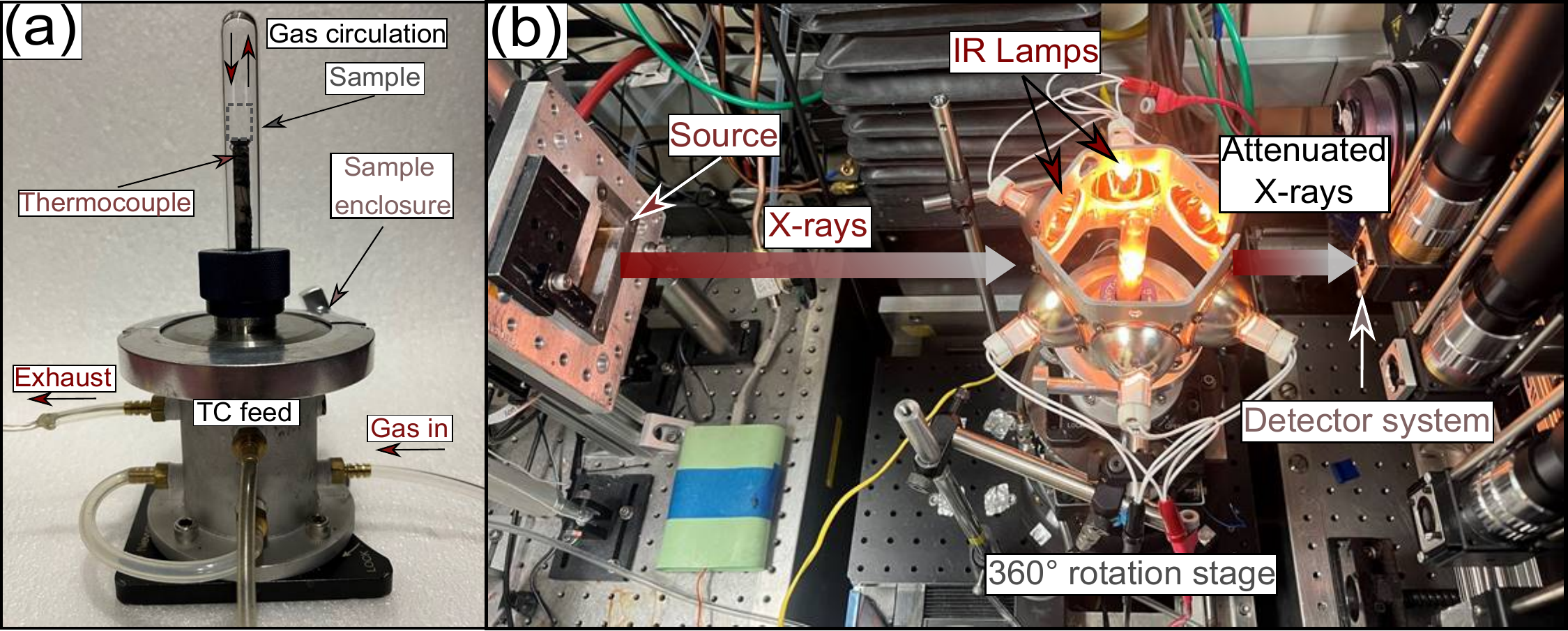}
	\caption{\emph{In situ} $\mu$-CT at the ALS. a) Controlled environment enclosure housing the sample. b) Sample enclosure installed in the focal spot of 6 IR heating lamps in the ALS $\mu$-CT hutch.}
	\label{fig:f0}
\end{figure}

The experiments are conducted at the Advanced Light Source (ALS) synchrotron facility at Lawrence Berkeley National Lab. Data is collected at the beamline 8.3.2, where \emph{in situ} $\mu$-CT can be performed using the high flux of X-rays directly from the synchrotron source, combined with environmental chambers to test materials under thermal, mechanical and chemical loads \cite{bale2013real, macdowell2016high}. The inner morphology of the material is evaluated during heating through continuous tomographies being generated every 90~seconds. This temporal resolution enables microstructural analysis beyond that of a typical lab-based computed tomography device. The RTV samples are contained in a sealed quartz-chamber as shown in Fig.~\ref{fig:f0}a, flushed with inert Argon gas constantly being cycled at 0.091 g/s to prevent oxidation from occurring at elevated temperatures. The sample enclosure is aligned in the beam path while also being surrounded by six confocally arranged infrared lamps aimed on the RTV cylinder (Fig. \ref{fig:f0}b) to increase temperature to 1000\C, similar to previous \emph{in situ} work \cite{maire2014,barnard2018,bale2012,haboub2014,macdowell2016}. The power supplied to the lamps is controlled such that a range of heating rates could be evaluated, this included: 54\C/min, 28\C/min, 14\C/min, and 7\C/min. Experiments are performed at temperatures ranging from 23\C (virgin material) to 1000\C (charred). A lens with a resolution of 3.25 $\mu$m/pixel is used for the high heating rates (54 and 28\C/min), and  a higher resolution 1.303 $\mu$m/pixel for the low heating rates (14 and 7\C/min) to further evaluate micro-pore evolution at smaller length scales. The samples are shaped to fit completely within the field of view (FOV) of the lens, \SI{3.3}{\milli\meter} diameter for the high heating rates and \SI{2}{\milli\meter} diameter for the low heating rates. Lastly, the smaller pores ($<$\SI{1}{\micro\meter}) below the resolution of the $\mu$-CT are qualitatively imaged with Scanning Electron Microscopy (SEM) using a FEI ESEM Quanta 450 FEG (FELMI ZFE, Gras, Austria).

\section{Results}
\begin{figure}[t!]
	\centering
	\begin{subfigure}{0.49\textwidth}
		\centering
		\includegraphics[width=\textwidth]{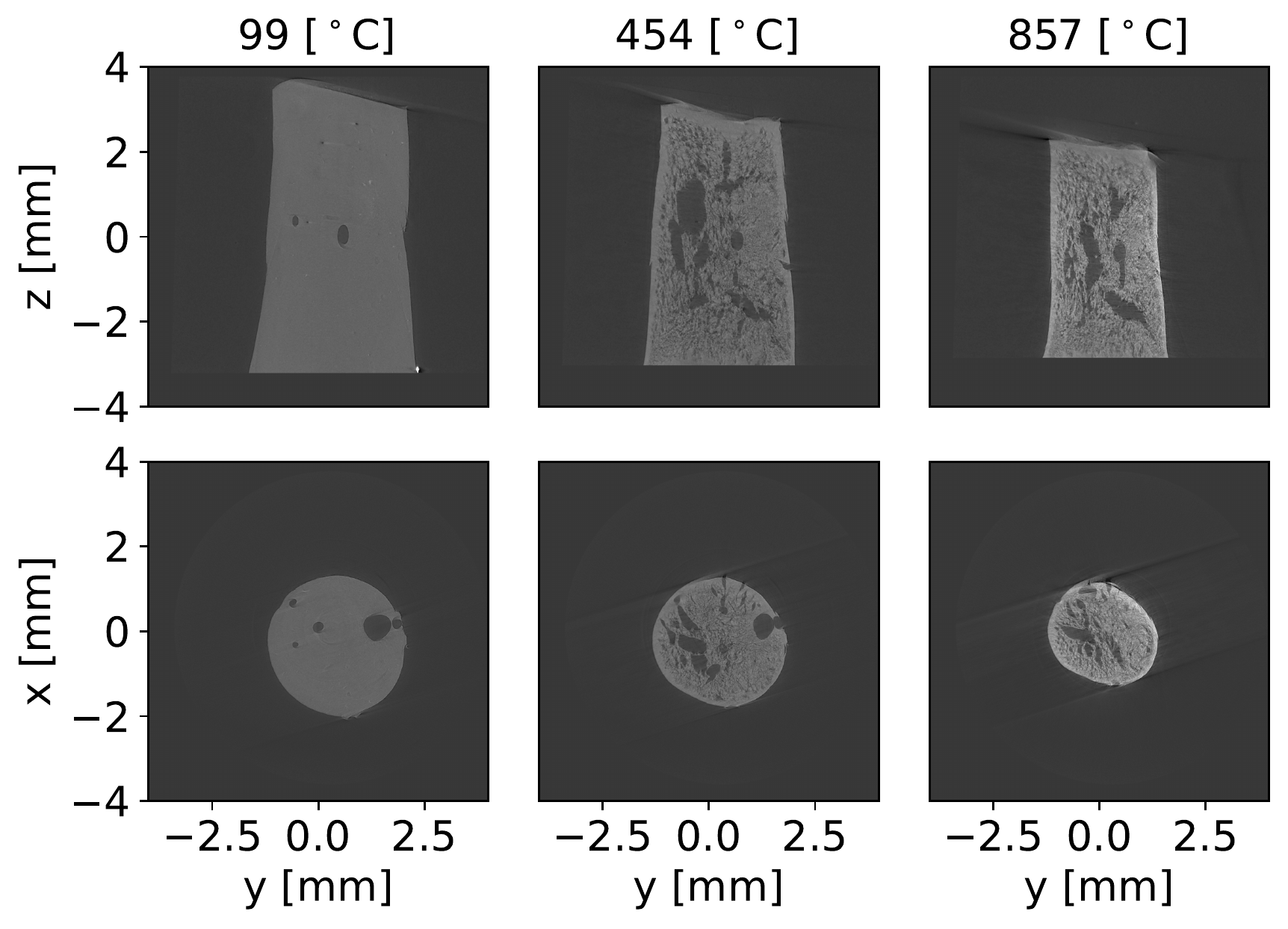}
		\caption{54\C/min}
		\label{fig:f3:11}
	\end{subfigure}
	\begin{subfigure}{0.49\textwidth}
		\centering
		\includegraphics[width=\textwidth]{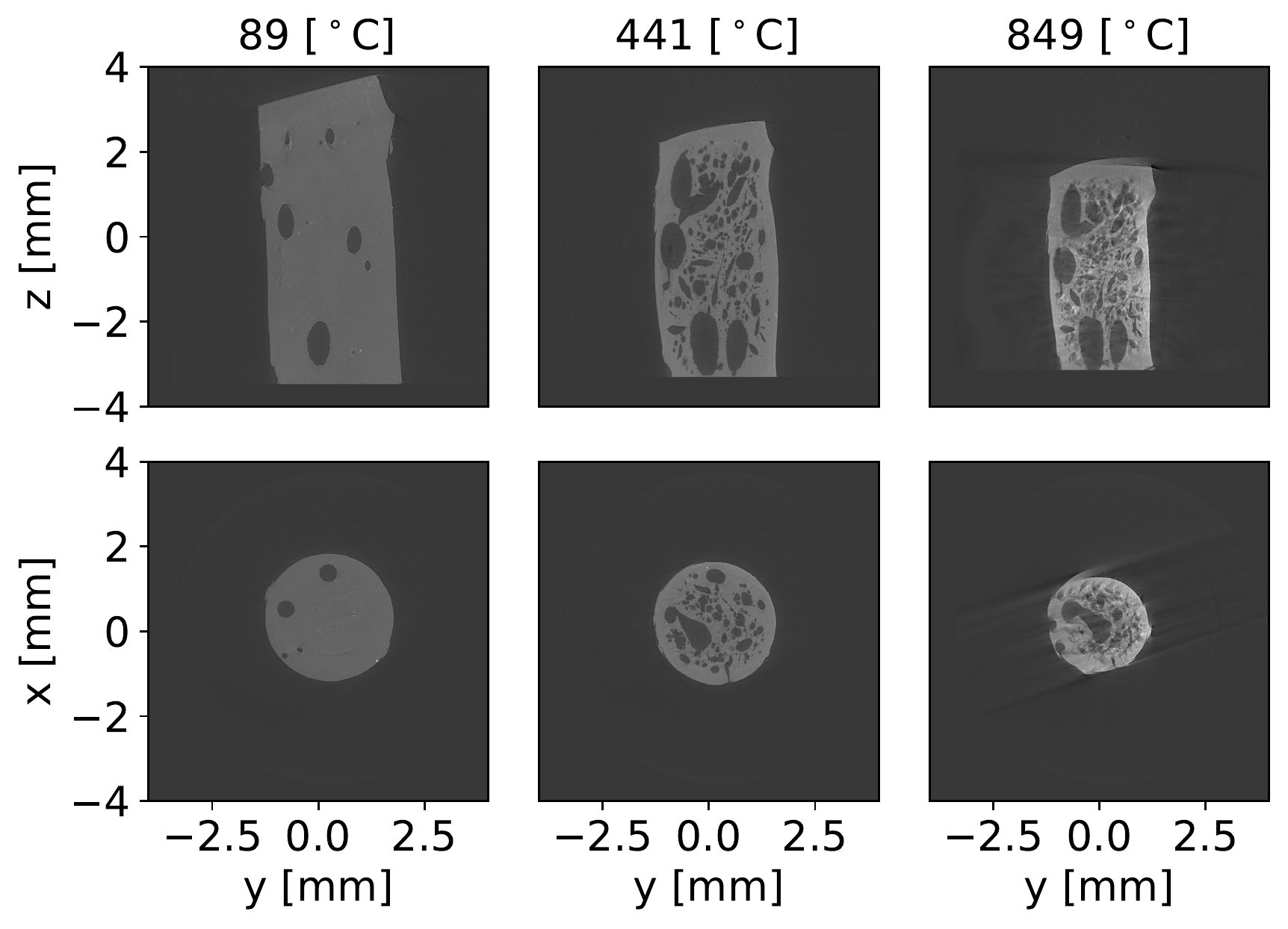}
		\caption{28\C/min}
		\label{fig:f3:12}
	\end{subfigure} \\
	\begin{subfigure}{0.49\textwidth}
		\centering
		\includegraphics[width=\textwidth]{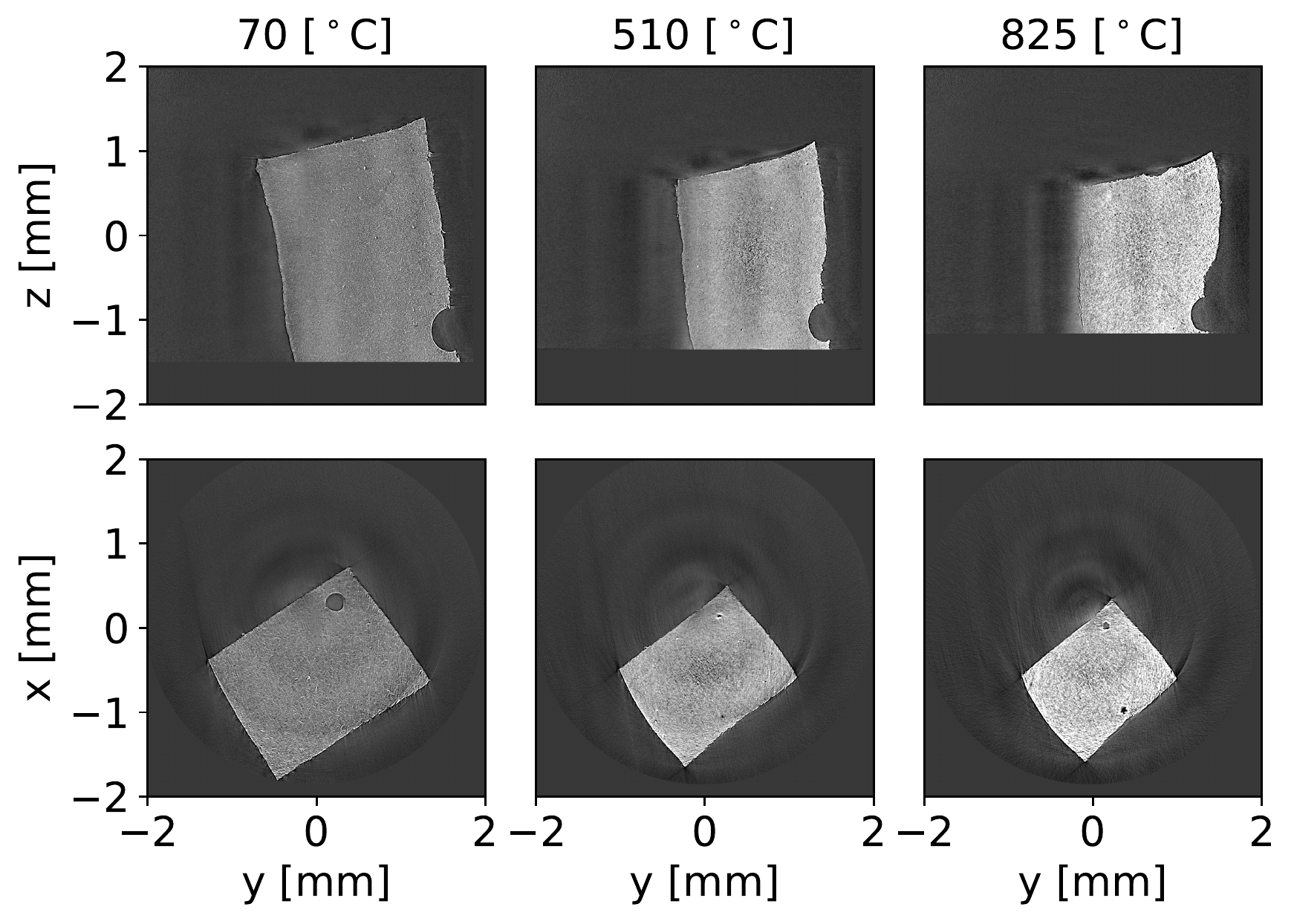}
		\caption{14\C/min}
		\label{fig:f3:31}	
	\end{subfigure} 
	\begin{subfigure}{0.49\textwidth}
		\centering
		\includegraphics[width=\textwidth]{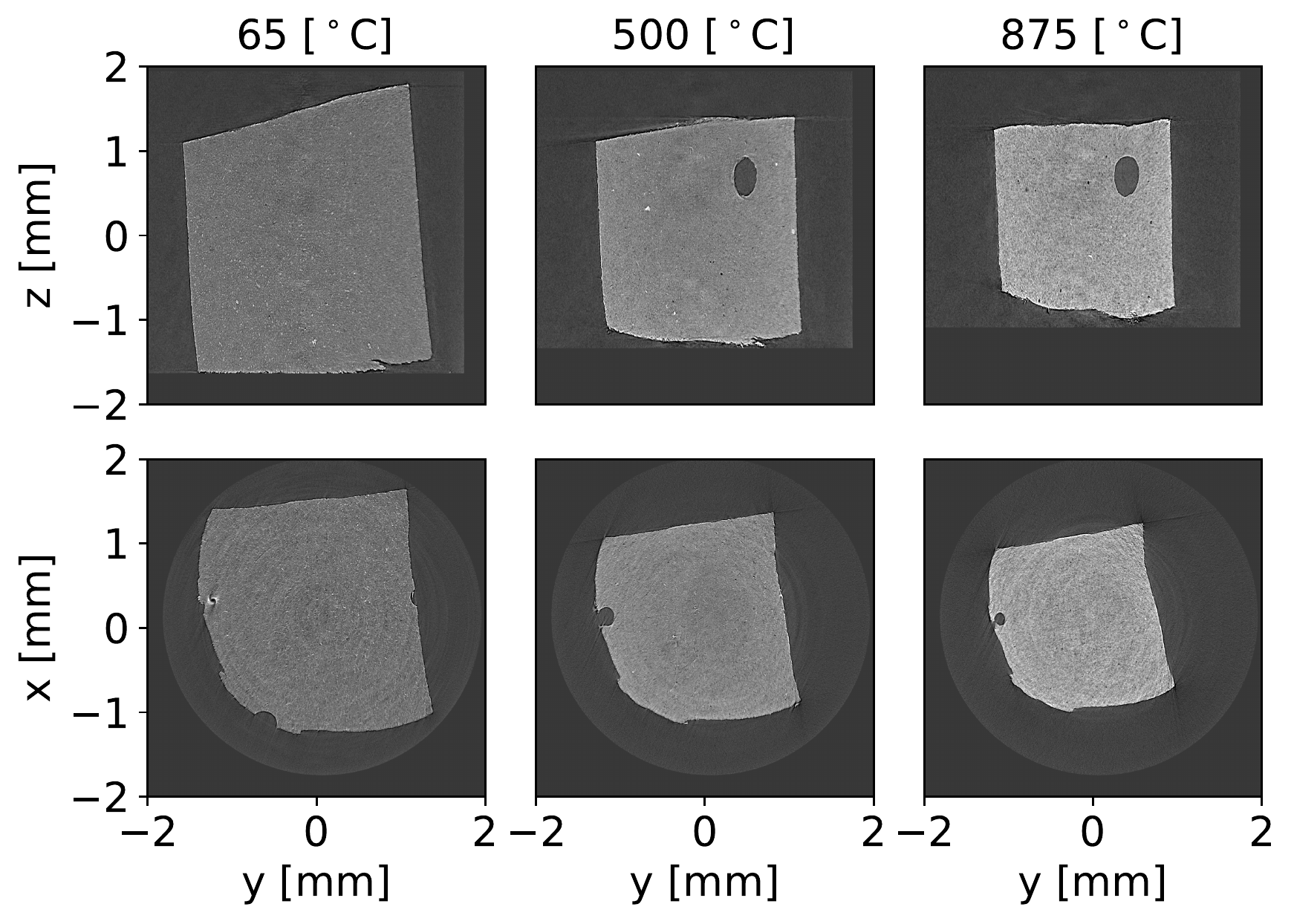}
		\caption{7\C/min}
		\label{fig:f3:32}
	\end{subfigure}
	\caption{Grayscale output from $\mu$-CT.} 
	\label{fig:f3}
\end{figure}


Results are presented in Fig. \ref{fig:f3} for the four heating rates investigated. Sample cross sections are shown from reconstructed $\mu$-CT datasets at increasing temperatures. Beginning with 54\C/min, Fig. \ref{fig:f3:11} shows the virgin sample with small voids formed that grow and interconnect into large porosities shown in its mid pyrolysis region (500\C). Along with the clearly visible pore network, small shear porosities (thin with high eccentricity) develop in the surrounding virgin region, further connecting in the late charring regime ($>$800\C). Next in Fig. \ref{fig:f3:12} at 28\C/min, the sample similarly begins with pores from curing that contribute in joining adjacent voids forming during peak pyrolysis. Seen in both Fig. \ref{fig:f3:11} and Fig. \ref{fig:f3:12}, while the sample generally experiences a reduction in height, swelling is observed from the outer circumference where the incident thermal radiation from the lamp heating occurs. This swelling behavior is clearly observed in both high heating rates (28 and 54\C/min) due to the rapid build up of internal pressure that causes a anisotropic distribution of new pores to form until the pyrolysis gases are liberated through open pores. 

For the low heating rates shown in Fig. \ref{fig:f3:31} and Fig. \ref{fig:f3:32}, smaller sample sizes are used to achieve higher spatial resolution while keeping the sample entirely in the FOV. Less initial defects (voids) are seen, but the same swelling and shrinking behavior is observed as the virgin material pyrolyzes to its charred state. Qualitatively, much less micro-pore development is observed with the lower heating rates (7 and 14\C/min), indicating a large dependence on heating rate. As further detailed in the subsequent quantitative analysis of the $\mu$-CT data, the tomographies are seen to get brighter in the solid phase, due to an increase in the X-ray attenuation of the sample. The increased X-ray absorption relates to the chemical structure transition from virgin RTV to char, and can be used as an estimate in material density shift. Additional video files of the RTV pyrolysis for all heating rates is provided in the supplementary material section. 


\begin{figure}[t!]
	\centering
    \includegraphics[width=\textwidth]{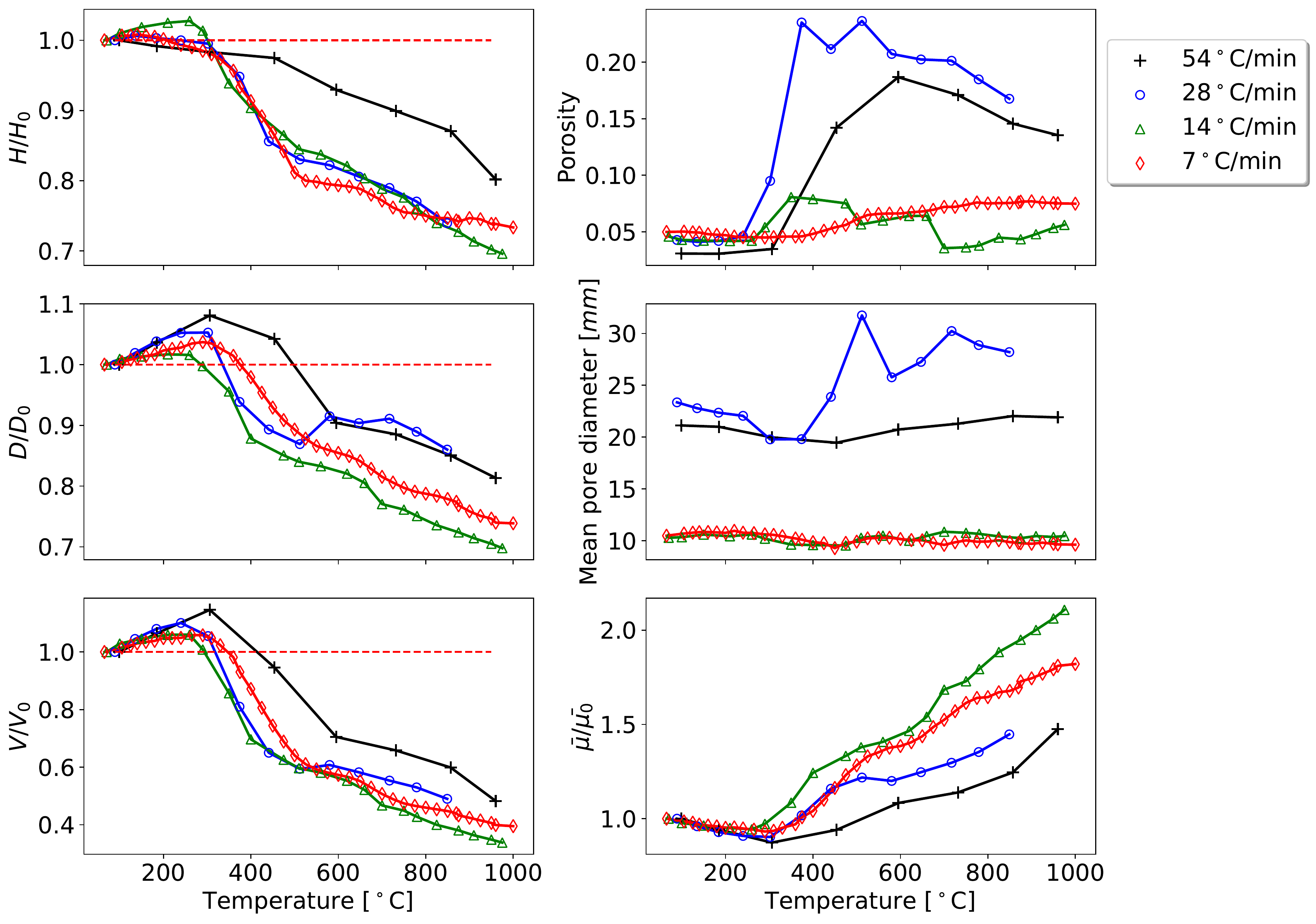}
	\caption{Quantitative analysis of RTV morphological evolution during pyrolysis.}
	\label{fig:f4}
\end{figure}

The quantitative behavior of RTV decomposition for the four heating rates is summarized in Fig. \ref{fig:f4}. 
First we examine the overall change in sample height, and we see that there is a percentage loss of 20-30\%. The higher heating rates tend to create a more anisotropically distributed pore network, contributing to a less-predictable change in the bulk volume as the pyrolysis progresses. 

The incident heat flux is applied radially to the sample, encapsulating a \SI{5}{\milli\meter} diameter sphere. As a result, the change in mean sample diameter is shown here to be similar for all samples. The initial solid swelling caused by internal gas pressure build-up is seen in the region leading to peak pyrolysis at 200-450\C. This swelling is followed by an  outgassing of pyrolysis byproducts and a shrinkage of the solid phase giving the reduction in diameter seen for the remainder of the pyrolysis process.

The result of the structural shifts in height and diameter are then fully realized by computing the change in solid volume during pyrolysis. Across all heating rates the sample is seen to swell at low temperatures ($\lesssim$400\C) during the initial phase of pyrolysis, as a result of gas build-up inside the material. Intumescence is found to be heating rate dependant with a volume increase up to nearly 15\% at the highest heating rate. An increase in heating rate is also observed to delay peak intumescence towards higher temperatures. At higher pyrolysis temperatures ($\gtrsim$400\C) there is a prominent material shrinkage that leads to up to 60\% volume reduction for the fully charred material, compared to the virgin sample. Shrinkage rate is higher in the 400-600\C range, where the bulk of the pyrolysis occurs \cite{oruganti2023} and decreases in the 600-1000\C range.

While the higher heating rates create larger, more interconnected porosities shown in Fig.~\ref{fig:f3}, the lower heating rates exhibit dissimilar behavior shown in the porosity measurements. We observed an increase in porosity up to nearly 20\% in the 400-600\C  pyrolysis region, with both 28\C/min and 54\C/min producing numerous pores that join neighboring voids until eventually reaching the surface. As the reaction progresses beyond peak pyrolysis gas release ($>$600\C), volumetric shrinkage begins to dominate in an overall reduction and collapse of pores within the samples. Similar behavior can be observed for the 14\C/min sample, as it experiences an increase in porosity through pyrolysis followed by a subsequent decrease thereafter. To be noted is the behavior of the slowest heating rate, 7\C/min, as porosity appears to monotonically increase during pyrolysis. This likely is the case as porosity is increasing on the nano-scale but is not resolved by the 1.303 $\mu$m/pixel resolution of the lens.

Porosity variation is further examined by the mean pore diameter. Shown to be largely unchanging for the two lower heating rates, they fluctuate slightly with the opening and closing of nano-porosities nearly undetectable by the resolution of the $\mu$-CT. Because the two higher heating rates begin with larger samples, they also have larger pre-existing pores that account for larger initial mean pore diameters. Where the 28\C/min generates more numerous disconnected porosities, the pores of the 54\C/min case coalescence with neighboring pores that do not shift dramatically in size nor shape.

Lastly, the change from initial material attenuation is plotted for all heating rates to show the change in relative density of the RTV. The change in attenuation was also qualitatively observed in Fig.~\ref{fig:f3}. All the RTV samples experience a small decrease in material attenuation leading up to peak pyrolysis (200-400\C) during volumetric swelling due to pressure build-up of the air trapped in the initial voids, followed by pyrolysis gas release, and subsequent cross-linking reactions that continue to increase the attenuation monotonically to char. This observed trend in attenuation is subject to further analysis by incorporating mass measurements to get predictions on density transition. 


\begin{figure}[H]
	\centering
    \includegraphics[width=\textwidth]{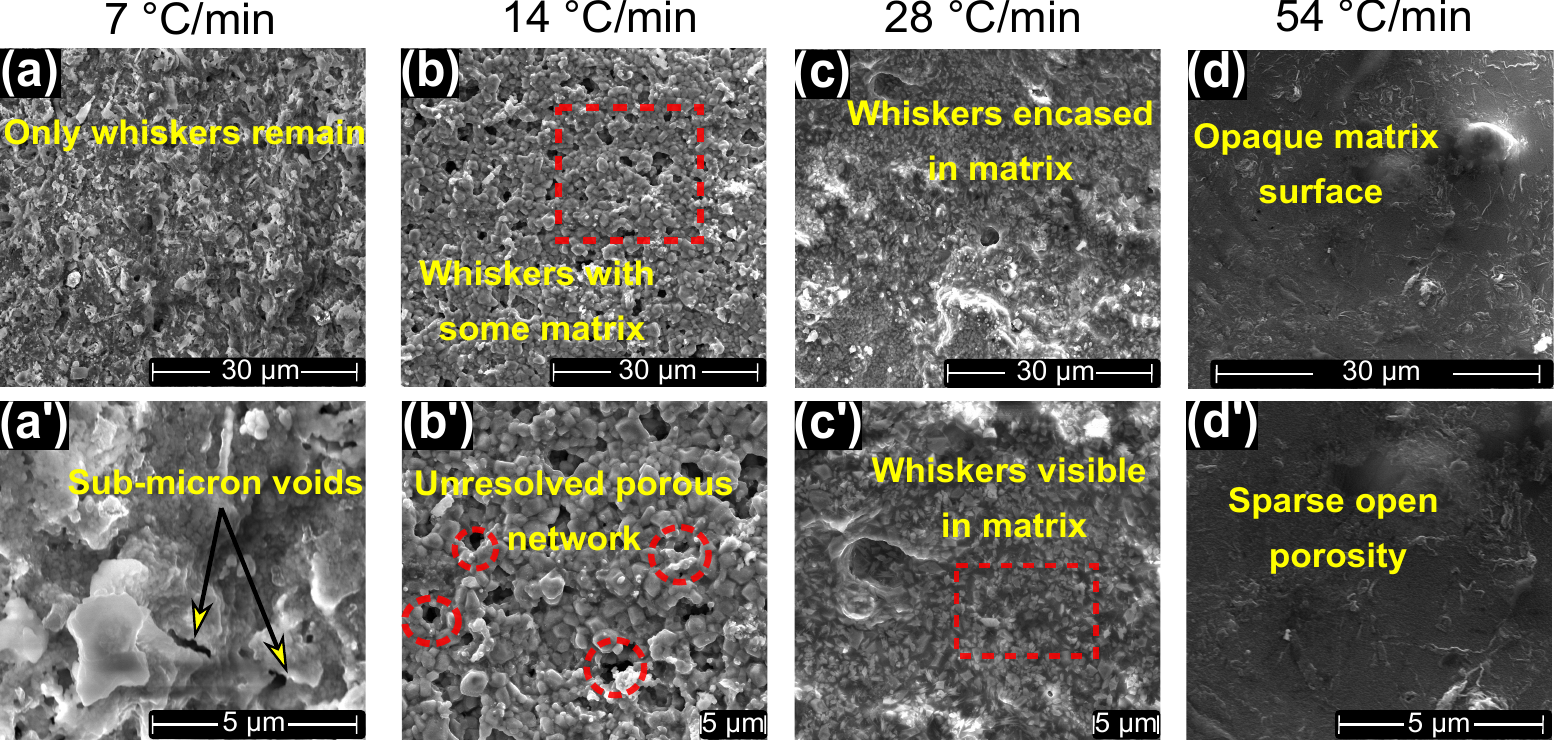}
	\caption{SEM of the 1000\C charred RTV samples.}
	\label{fig:f5}
\end{figure}

SEM images for the tested materials are shown in Fig.~\ref{fig:f5}. The RTV examined is a two-phase material with a matrix engulfing smaller solid particles deemed whiskers. Previous energy dispersive electron spectroscopy described the whiskers and remaining matrix phase being a slurry of C, O, Si, and Fe \cite{oruganti2023,oruganti2022}. The iron in the RTV matrix transitions from a red Hematite (Fe\textsubscript{2}O\textsubscript{3}) to the black Magnetite (Fe\textsubscript{3}O\textsubscript{4}), with charred Si, C, and SiO\textsubscript{2} matrix when fully-charred; decomposition reactions and the release of volatile compounds of O-Si-C chains account for the mass loss in the RTV \cite{oruganti2023}. This is first visualized in Fig. \ref{fig:f5}a-a' for the 7\C/min case. At low heating rates the resulting microstructure resembles the observations of past TGA investigations at 10\C/min \cite{oruganti2022,oruganti2023}. More of the matrix material is pyrolyzed and coalesces with the whiskers, exposing porosities at the sub-micron scale not resolved by the $\mu$-CT. The same is observed to a lesser degree for the 14\C/min sample, as there is evidence of the whiskers material still exposed. After the final tomography is taken at 1000\C, the heating is immediately shut-off and the sample is cooled to room-temperature, quenching any further pyrolysis reaction. Therefore, at the highest heating rates pyrolysis remains incomplete. Moving to 28\C/min, it is seen that the matrix phase has not fully pyrolyzed, with exposed whisker material visible in \ref{fig:f5}c'. This trend is further emphasized by the 54\C/min case, with uniform outgassing leaving behind a more opaque surface with few nano-porosities that would have likely formed given more time at high temperature. Evidence of incomplete pyrolysis of the material is also seen in the plot of X-ray attenuation in Fig.~\ref{fig:f3}, with the lower attenuation values for the high heating rates, indicating pyrolysis not being complete.  

\begin{figure}[H]
	\centering
    \includegraphics[width=\textwidth]{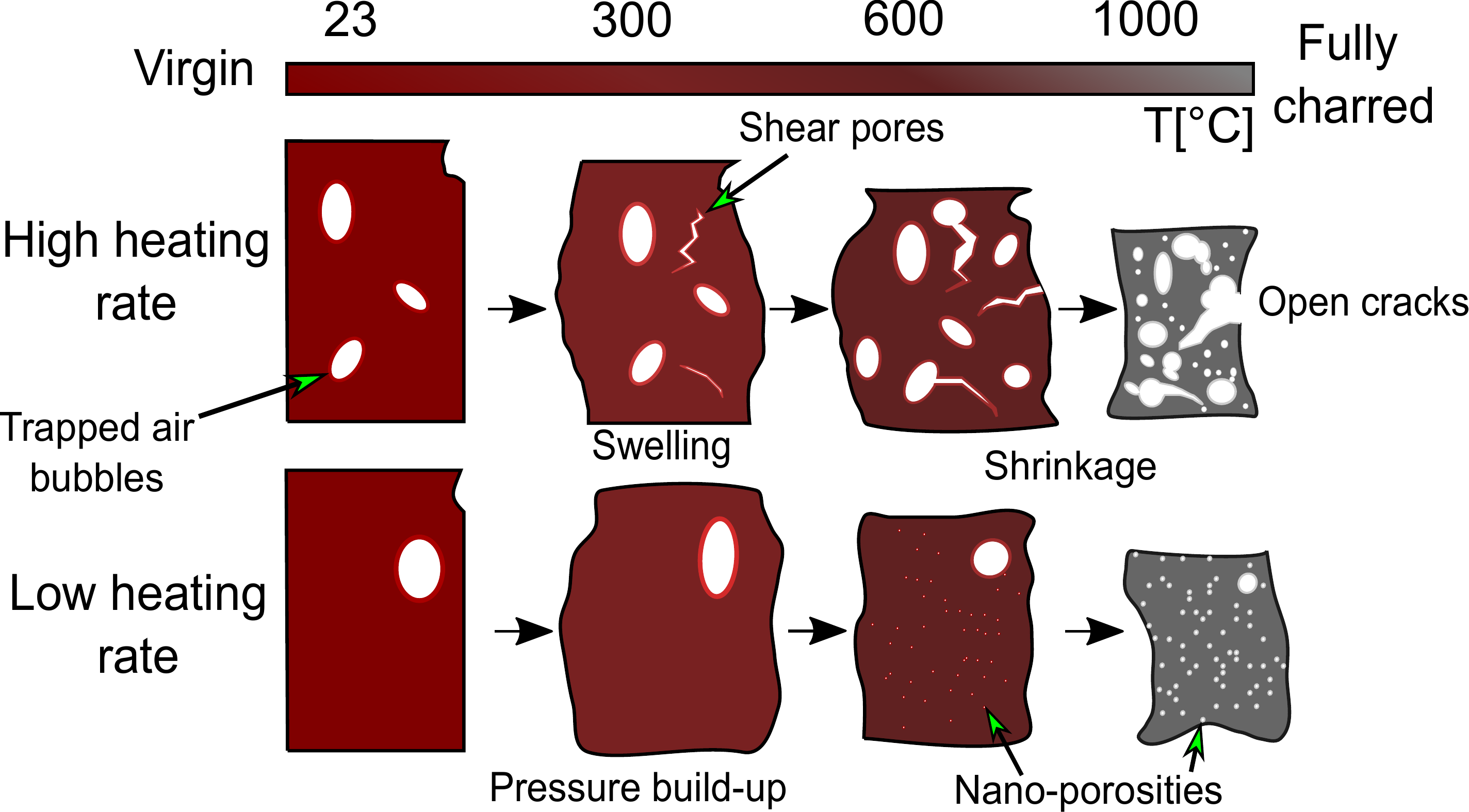}
	\caption{Decomposition model of RTV.}
	\label{fig:f2}
\end{figure}

A generalized graphic summarizing the key observations of the pyrolysis of RTV between high (28 and 54\C/min) and low (7 and 14\C/min) heating rates is shown in Fig.~\ref{fig:f2}. Both non-pristine samples begin similarly with slight volumetric expansion, although the higher heating rate experiences a larger degree of new porosity development, where the low heating rates generally just expand the pre-existing pores. The pyrolysis continues and the high heating rate experiences large internal pressure gradients causing the development of anisotropically distributed pores that interconnect with the pre-existing pores. The low heating develops nano-pores as the pyrolysis process nears closer to completion. Lastly, both samples experience large volumetric shrinkage as mass is lost from pyrolysis gas escaping, and the overall porosity decreases as the solid volume collapses.  


\section{Conclusion}
In conclusion, this study is a critical contribution to informing micro- and meso-scale behavior of material systems that utilize RTV in high temperature environments. Examined is a quantitative analysis of the micro-structural development of RTV under a range of relevant pyrolysis heating rates. The results show stark differences from a higher heating rate ($>$28\C/min) to a low rate (7-14\C/min) in porosity formation and morphology. Bulk material swelling and shrinking is also observed and is relatively similar across all heating rates, critical when evaluating mechanical performance of the material system. The continuation of this work will investigate the densification of this material utilizing the attenuation measurements, and conduct a similar set of experiments on RTV samples that are defect-free from manufacturing. This information will be fed directly to multi-physics reentry and fire codes to evaluate for relevant effective transport properties. 



\section*{Acknowledgements}
This work is supported by a NASA Space Technology Graduate Research Opportunities Award Grant No: 80NSSCC22K1192 and 80NSSC21K1117. This research used resources of the Advanced Light Source, which is a DOE Office of Science User Facility under contract no. DE-AC02-05CH11231. This material is based upon work supported by the U.S. Department of Energy, Office of Science, Office of Workforce Development for Teachers and Scientists, Office of Science Graduate Student Research (SCGSR) program. The SCGSR program is administered by the Oak Ridge Institute for Science and Education for the DOE under contract number DE‐SC0014664. Work was also done in part at the Beckman Institute Microscopy Suite at the University of Illinois at Urbana-Champaign. We would also like to thank the support of Dula Parkinson and Harold Barnard at the ALS for their guidance in experiments. 

\bibliographystyle{unsrt}  
\bibliography{references}

\end{document}